\documentclass[a4paper,11pt]{article}
\pdfoutput=1 
\usepackage[utf8]{inputenc}
\usepackage{jheppub}
\usepackage{graphicx,xcolor,amsmath,amssymb,fancyhdr,siunitx,mathtools}
\usepackage[numbers]{natbib}
\usepackage[titletoc]{appendix}
\numberwithin{equation}{section}

\usepackage{cleveref}

\def\dslash{\not{\hbox{\kern-2pt $\partial$}}}
\def\Dslash{\not{\hbox{\kern-4pt $D$}}}
\def\Oslash{\not{\hbox{\kern-4pt $O$}}}
\def\Qslash{\not{\hbox{\kern-4pt $Q$}}}
\def\pslash{\not{\hbox{\kern-2.3pt $p$}}}
\def\kslash{\not{\hbox{\kern-2.3pt $k$}}}
\def\qslash{\not{\hbox{\kern-2.3pt $q$}}}
\def\svslash{\not{\hbox{\kern-2.3pt $sv$}}}
\usepackage{verbatim}

\newcommand{\md}[1]{\textcolor{blue}{[MD: #1]}}

\title{The Problem of Axion Quality:  A Low Energy Effective Action Perspective}

\author{Michael Dine,}

\emailAdd{mdine@ucsc.edu}

\affiliation{
    Santa Cruz Institute for Particle Physics and Department of Physics,\\University of California, Santa Cruz,\\
    Santa Cruz, CA, USA
}

\abstract{Any would-be Peccei-Quinn (PQ) symmetry is vulnerable to various types of explicit breaking.  It has long been recognized that these can disrupt the axion
solution to the strong CP problem.  There have also been suggestions that, under certain circumstances, these can lead to a surprisingly large mass for the axion.
Two types of corrections
to this computation have been widely considered:  higher dimension symmetry breaking operators, in theories where there is a complex field responsible for the spontaneous
breaking of the PQ symmetry, and small instantons.  Motivated by situations where small instantons dominate the $\theta$ potential of non-abelian gauge theories, we formulate the question of axion quality in terms of constraints on a Wilsonian effective action at scales somewhat above the scale of QCD.  In this language, the standard axion mass computation assumes a nearly exact PQ symmetry in this action.
The higher dimension operators and/or small instantons represent symmetry breaking terms in the Wilsonian action.  This picture permits a uniform treatment of the problem
of {\it axion quality}.   If one assumes order one CP violation at high energies, then solving the strong CP problem constrains the axion potential terms in this Wilsonian action; in particular, the axion mass
must be extremely close to its ``standard" value.}

\begin{document}
\maketitle

\section{Introduction and Overview}

The axion solution to the strong CP problem relies on the existence of a global, Peccei-Quinn (PQ) symmetry of {\it high quality}.  As we don't expect the underlying theory of nature to exhibit exact global symmetries,
this requires some explanation.  In the literature, at least two sources of symmetry violation have been considered:
\begin{enumerate}
\item  In situations where the effective theory, at the Peccei-Quinn scale, contains a field, $\Phi$, responsible for spontaneous breaking of the PQ symmetry, high dimension operators involving $\Phi$ can explicitly
break the symmetry\cite{mrk}.  Typically, operators of very high dimension must be suppressed.  This can be achieved, for example, with intricate discrete symmetries\cite{dinewittensymposium}.
\item  QCD itself violates the symmetry.  In situations where the QCD $\beta$ function is small at high energies, small scale instantons can make substantial contributions to the axion potential, possibly dwarfing the standard result.  This was pointed out some time ago in \cite{holdompeskin}, and reiterated
more recently in \cite{shifmanetal, hookheavyaxions,gherghetta,kitanosmallinstantons,grandcolor}.
\end{enumerate}
But as pointed out in \cite{dineseibergsmallinstantons}, such small instanton contributions are sensitive to unknown short distance CP violating effects,
raising the question of whether unknown, and possibly uncontrolled, non-perturbative string theory effects might spoil
the PQ solution.  String theory, which motivated that study, is remarkable, in fact, in providing approximate PQ symmetries at weak coupling.  But how large the breaking of the symmetry might be is an open question, for which
the instanton analysis, at best, suggests lower bounds.
 
 It is worth articulating what distinguishes small and large instantons.  A standard way to compute the axion potential in QCD is to study an effective theory
 at scales below the scale of chiral symmetry breaking, which includes eight Goldstone bosons and the axion. In addition to the eight axial $SU(3)$ currents and the corresponding Goldstone bosons, one can then define a ninth
 current which is free of anomalies.   This current is not conserved already classically, with divergence proportional to light quark masses. Matrix elements of this current
 are studied in the non-linear sigma model.  Thinking of this as a Wilsonian theory with an upper cutoff $M \sim 1 ~{\rm GeV}$, this procedure accounts for
 non-perturbative effects such as instantons at scales below $M$.  Small instantons and other short distance non-perturbative effects would yield PQ symmetry
 violating terms in the Wilsonian action, and these affects must be accounted for separately.

In this note, in addition to providing a unified view of these problems, we ask the extent to which these types of contributions to the Wilsonian action endanger the PQ solution of the strong CP problem.  If one assumes
order one CP violation at the relevant energy scales, we will see that the axion mass must be extremely close to its value in the standard computation.  The question of higher dimension operators is well studied\cite{mrk} and the results can be understood in this language.  For small
instantons, this translates into constraints on the $\beta$ function.  Interesting models, such as minimal gauge mediation (and certainly non-minimal gauge mediation) are susceptible to large corrections.
Indeed, in such cases, small instantons are potentially disruptive of supersymmetry breaking itself.

In the next section, we discuss the Wilsonian effective action.   In section \ref{higherdimensionoperators}, we review the issues associated with higher dimension operators in theories in which PQ symmetry breaking
is driven by a single (or small number of) complex fields.  In section \ref{smallinstantons}, we review the potential problems associated with small instantons, noting that, quite generally, in non-supersymmetric theories, if the leading term in the QCD $\beta$ function is less than four, then generically small instantons dominate, and high energy CP violation endangers the PQ solution of the 
strong CP problem.   In section \ref{supersymmetry}, we consider the question of small instantons in supersymmetric theories.  Here, on the one hand, the $\beta$ function, for a given number of generations, is smaller; on the other hand, the need for supersymmetry-breaking insertions softens the high energy behavior.  We review, as noted in \cite{dineseibergsmallinstantons}, the argument that the axion potential is first order in the gluino mass.  We explain why, in cases where they are the dominant contribution to the axion potential, small instantons are potentially the principle source of supersymmetry breaking. 
Noting, as discussed in \cite{dineseibergsmallinstantons}, the way in which small instanton effects may be cut off in grand unified theories above the unification scale, in section  \ref{unknown}, we speculate
on possible behaviors of instanton computations in ``ultraviolet complete" theories such as string theory.
 In section \ref{conclusions}, we consider the implications of these results for
particular cases:  non-supersymmetric theories with more than four generations of quarks and leptons, supersymmetric theories with additional quarks and leptons, such as required in gauge mediated models,
and other situations.  We stress that, quite generally, corrections to the standard formula for the axion mass must be extremely small if the axion is to solve the strong CP problem without strong assumptions about CP violation at short distances.

\section{The Axion Effective Action}
\label{wilsonianaction}

We consider, here, a Wilsonian effective action for QCD with an axion, defined at a scale above the scale of QCD, but well below the weak scale. 
We assume an approximate PQ symmetry, and write an effective action for the axion field, analogous to the non-linear effective action for the pseudoscalar mesons.  The action is built
out of an object with nice transformation properties under the Peccei-Quinn symmetry:
\begin{equation}
\Phi= f_a e^{i a \over f_a}.
\end{equation}
The action includes derivative terms, which respect the PQ symmetry.  It includes also PQ violating potential terms, as well as derivative terms which violate the symmetry.  If these terms are small,
the computation of the axion mass is the standard one, plus perturbations.

More precisely, the effective action has the structure:
\begin{equation}
{\cal L} = -{1 \over 4g^2} F_{\mu \nu}^2 +  {a(x)\over 16 f_a \pi^2} F \tilde F + {\rm quark~kinetic~and~mass~terms} 
\label{wilsonianaction}
\end{equation}
$$~~~~ + {1 \over 2} \vert \partial_\mu \Phi\vert ^2 + {\rm symmetry~violating~derivative~terms}
+\sum_i \Lambda_i^{4-n_i}  \Phi^n_i + {\rm c.c.} + \dots$$
Here the $\Lambda_i$ terms represent symmetry breaking effects, and the parameters $\Lambda_i$ are themselves, in general, complex.  We have defined $a$ through its coupling to $F \tilde F$.

If the potential terms are small, in the standard way, we can calculate the axion potential from this action.  Recall that this ``Standard Computation" proceeds by first modifying the axion
current defining a non-anomalous,
non-conserved current; the corresponding symmetry is classically violated by light quark mass terms.  Then
the quark mass terms are proportional to $e^{i A{a \over f_a} + i {\vec \pi \cdot \vec \tau\over 2 f_\pi}}$\footnote{This can be seen by rewriting these terms in terms of the degrees of freedom
of the non-linear sigma model, and requiring that, in the ground state, there be no tadpole for the pions or the axion.}.  The leading axion mass may be calculated in a straightforward way.

We can then treat the $\Lambda_i$ terms as perturbations.  This Wilsonian setup allows a uniform treatment of:
\begin{enumerate}
\item  Higher dimension, PQ symmetry violating operators originating from an effective action at scales well above QCD scales\cite{mrk}.
\item  Small instantons
\item  Unknown non-perturbative effects, e.g. in string theory.
\end{enumerate}

Let's consider each of these in turn.

\section{Higher Dimension Symmetry Violating Operators}
\label{higherdimensionoperators}

Suppose we have a theory with a complex field, $\phi$, $\langle \phi \rangle = f_a$, transforming under the PQ symmetry, which acquires an expectation
value of order $f_a$, and which creates a light axion and a massive field.   Then it has long been known\cite{mrk} that symmetry violating operators
of the form
\begin{equation}
{\cal L}_{symmetry-breaking} = {1 \over \Lambda^{n-4} } \phi^n + {\rm c..c}
\end{equation}
endanger the PQ solution to the strong CP problem, even if $\Lambda \sim M_p$. Indeed, below the mass of ${\rm Re}~\phi$, htese operators have precisely the form of our symmetry violating operators in the Wilsonian
effective action of equation \ref{wilsonianaction}.  Achieving the required {\it quality} of the Peccei-Quinn symmetry requires suppressing operators up to $n \sim 12$.  If this is to be
achieved through, say, discrete symmetries, these must be rather intricate\cite{dinewittensymposium}.

\section{Small Instantons in Non-Supersymmetric Theories}
\label{smallinstantons}

The possibility that small instantons might enhance the axion mass was raised in \cite{holdompeskin}.
In \cite{dineseibergsmallinstantons}, motivated by then popular string theory constructions, theories were considered where, at high energy scales, the QCD beta function is, at leading order, small. If one assumes that it is possible to tie together all of the fermion zero modes, with no additional chiral suppression, then, writing the leading
order term in the $\beta$ function, $b_0$:
\begin{equation}
\beta(g) = - b_0 {g^3 \over 16 \pi^2} 
\end{equation}
one has, for the instanton contribution to the axion potential at small $\rho$:
 \begin{equation}
K
e^{i {a \over f_a}  +i \delta}\int {d\rho \over \rho^5} e^{-{8 \pi^2 \over g^2(\rho)}} + {\rm c.c.} = e^{i {a \over f_a}  +i \delta}\int {d\rho \over \rho^5} (\Lambda \rho)^{b_0}
\end{equation}
for some constant $K$.
For small enough $b_0$, 
\begin{equation} 
b_0 \le 4
\end{equation}
small instantons dominate.  In general, the phase, $\delta$, depends on unknown high energy physics.  To make sense of this expression, we must assume that the scale size integral is cut off by some 
additional dynamics at some small $\rho= \rho_0$.  In this case, again, the action is of the general form we have described above.

To be specific, following \cite{dineseibergsmallinstantons},  we can consider an $SU(5)$ grand unified theory.  With three generations and without additional colored scalars or other fields,
the $\beta$ functions for $SU(3)$ and $SU(2)$, below the unification scale, are:
\begin{equation}
b_0^{(3)} = 7; b_0^{(2)} = {10 \over 3}.
\end{equation}
For $SU(3)$, the $\beta$-function is not small enough that small instantons dominate.  For $SU(2)$ they do, but assuming coupling unification, $SU(2)$ instantons are still highly suppressed at the unification
scale.  Above that scale, the $b_0$ of the full $SU(5)$ theory is $42/3$, and instanton effects die out rapidly.

Things are different if there are thresholds for additional fields, say scalars transforming under $SU(3) \times SU(2) \times U(1)$, slightly above the weak scale,.  For example, if there are several
additional scalar $SU(3)$ triplets, such that 
\begin{equation}
b_0^{(3)} = {2},
\end{equation}
small instantons dominate up to the unification scale.  One can close up the instanton with insertions of the Higgs field; there might be additional possibilities to
obtain a non-zero axion potential involving new physics at very short distances.
From the Higgs field, the result is proportional to
\begin{equation}
M_{GUT}^4 e^{-{8 \pi^2 \over g^2(M_{GUT})}} {\rm det} (y_U ) {\rm det} (y_D) 
\end{equation}
Taking $M_{GUT}= 10^{16} {\rm GeV}$, this is orders of magnitude larger than typical QCD energy scales, $m_\pi^2 f_\pi^2$, even allowing possible substantial
numerical suppression.  As a result, the QCD vacuum angle is at least as large as the phase of the determinant of the CKM matrix, of order $10^{-5}$.  It might be larger
if there are additional CP violating effects at scales near the GUT scale.

It is perhaps worth recalling how one computes the effective $\theta$, ${\langle a \rangle \over f_a}$ assuming small instantons are suppressed.
At low energies, including only dimension four terms in the QCD lagrangian, the theory has a symmetry of CP $+ ~\theta \rightarrow -\theta$.
So before including higher dimension operators, the minimum of the potential for $\theta$ is at $\theta=0$.   Including higher dimension, CP violating operators, the
minimum shifts, by amount of order:
\begin{equation}
\theta_{min} = {\Lambda_{QCD}^2 \over M^2} \alpha.
\end{equation}
Here we have assumed a dimension six operator, scaling with $1/M^2$; $\alpha$ is a CP violating phase.  In the Standard Model, we would expect 
\begin{equation}
\alpha \sim {\rm Im} {\rm det} (Y_U)
 {\rm det} (Y_D).
 \end{equation}  The CP violating phase, in this case, is of order the phase of the determinant of the KM matrix, roughly $10^{-5}$.  So we see that, when small instantons are important, they dominate and lead to a far larger phase, even though the underlying CP violating
 phases
 have a similar origin.  This suggests that aligning the minimum of the $\theta$ potential associated with small instantons with that from low energy QCD
 is difficult.


\section{Small Instantons in Supersymmetric Theories}
\label{supersymmetry}

In the supersymmetric case, there are new features.  In particular, at energy scales much above the supersymmetry breaking scale (instanton scale sizes correspondingly smaller) the axion ($\theta$) potential
will be suppressed by powers of some supersymmetry breaking scale.  In \cite{dineseibergsmallinstantons} it was argued that one should be able to tie together all of the instanton zero modes by including a single
insertion of $m_\lambda$ (as well as suitable Yukawa couplings).  So now the axion ($\theta$) potential behaves as:
\begin{equation}
K^\prime
e^{i {a \over f_a}  +i \delta}\int m_\lambda {d\rho \over \rho^4} e^{-{8 \pi^2 \over g^2(\rho)}} + {\rm c.c.} = K^\prime e^{i {a \over f_a } +i \delta}m_\lambda\int {d\rho \over \rho^4} (\Lambda \rho)^{b_0}
\end{equation}
for some constant $K^\prime$.  In this expression, small instantons dominate if $b_0 \le 3$.  In the simplest $SU(5)$ model with three generations, $b_0 = 3$, so we are precisely at the boundary.  If there are additional chiral
multiplets in the $3$ and $\bar 3$, as in the simplest gauge mediated models, for example, we will be in a regime of small instanton domination, and the PQ solution of the strong CP problem
is vulnerable to unknown high energy CP-violating effects.  If, instead, instanton effects are second order in $m_\lambda$, things are slightly better.

It is interesting to consider the case that instanton effects are linear in $m_\lambda$ from another point of view. Thinking of $m_\lambda$ as linear in a supersymmetry breaking spurion, the potential
due to instantons is {\it linear} in the spurion, as would be the case for the leading term in an O'Raifeartaigh-like model.   If small instantons dominate, there is the potential for this linear term
to be the dominant source of supersymmetry breaking.  To gain some feeling for the potential problems, consider the possibility that there is an underlying gauge mediated theory with a single generation of vectorlike
messenger chiral fields. In that case, $b_0^{(3)} = -1$, and, assuming small instantons are cut off at $M_{GUT}$, the linear term in $m_\lambda$ is of order $M_{GUT}^3 \left ({M_{GUT} \over \Lambda_{QCD}} \right )$.
This is potentially a huge enhancement, though one also expects significant suppression by powers of Yukawa couplings and possible numerical factors.

\section{Unknown Non-Perturbative Effects}
\label{unknown}

If we consider the sorts of axions which arise in string theory, we might expect that PQ violating effects could arise from sources beyond those so far considered.  Non-perturbative effects in these theories
are not well understood, and one might expect effects at high energies larger than those anticipated from instantons.  The instanton effects we have described so far are presumably just a lower bound.  We have seen, for example, how the unification scale can act
as a cutoff on small scale QCD instantons.  The cutoff in a string theory might operate differently.  Factors of a few change in the effective cutoff could significantly
change the size of the result.   Lacking such an understanding, we have to acknowledge that the successful implementation of the PQ solution to the strong CP problem relies on strong assumptions about very short distance physicso.

\section{Conclusion:  Possible Implications of Small Instantons for Interesting Models}
\label{conclusions}

Simply allowing additional colored fermion or boson fields in the Standard Model, we have seen, could substantially enhance the role of small instantons.  Supersymmetric
theories are even more susceptible to such enhancements.  We have noted at least one well motivated model where small instantons would play a role:  theories of gauge
mediated supersymmetry breaking.

\subsection{Heavy Axions}

More generally, over the years and particularly quite recently, there has been some interest in solutions to the strong CP problem where the axion might be heavy; a partial list includes \cite{holdompeskin,shifmanetal, hookheavyaxions,gherghetta,kitanosmallinstantons,grandcolor}.  From the perspective of the effective action
for the axion which we have advocated in this paper, such heavy axions are problematic.  Again, by assumption, the action has an approximate $U(1)$ symmetry,
and the axion is a compact field.  So the effective action is a sum of terms of the form:
\begin{equation}
V(a) = {m_\pi^2 f_\pi^2} \cos({a \over f_a})+ \sum \Lambda_i^4 \cos({n_i a \over f_a} + \alpha_i)
\end{equation}
Here we have used an old argument of Weinberg's to write the leading QCD effects in terms of an even function of the axion field ($\theta$):   the low energy theory has a symmetry which is conventional CP accompanied by a reversal of the sign of $\theta$ ($a \over f_a$).

If $\Lambda_i$ for a particular $i$ is the dominant term here, then we have 
\begin{equation}
{\delta m_a^2 \over m_a^2} = {\Lambda_i^4 \over m_\pi^2 f_\pi^2}
\end{equation}
The effective $\theta$ is
\begin{equation}
\langle {a \over f_a} \rangle = \alpha_i {\delta m_a^2 \over m_a^2} 
\end{equation}
If the phase $\alpha_i$ is of order one, then we require:
\begin{equation}
{\delta m_a^2 \over m_a^2}  < 10^{-10}.
\end{equation} 
In other words, we can only have an enhanced axion mass if new sources of CP violation, connected with this mass, are small.  We have seen this in various contexts in this note:  higher dimension operators, small instantons.

\subsection{Gauge Mediated Supersymmetry Breaking}

We have noted that gauge mediated supersymmetry breaking, where $b_0^{(3)}\le 2 $ is particularly prone to domination by small instantons.  There is the possibility that such instantons in fact are the dominant source of SUSY breaking.  This may pose an obstacle to the implementation of the PQ solution of the strong CP problem in such
models, and possibly even an obstruction in some cases to the construction of such models. It is possible that in some models one can't close up the zero modes without multiple supersymmetry breaking insertions, thus suppressing small instantons.  This is a subject worthy of further investigation. 

\subsection{Pitfalls for Model Builders -- and Nature -- To Avoid}

We can summarize by saying that we require of nature, if we are to successfully implement the PQ solution of the strong CP problem
\begin{enumerate}
\item Before considering non-perturbative QCD effects, a high quality PQ symmetry.
\item  Non-perturbative QCD effects must not be larger than $10^{-10}$ of the axion mass-squared.
\end{enumerate}

By definition, we don't have direct experimental knowledge, at present, of these high energy effects. So the PQ mechanism to solve the strong CP problem is contingent on features of high energy physics
which are currently unknown.  Said another way, discovery of a Weinberg-Wilczek axion would set strong limits on possible ultraviolet physics.

\section*{Acknowledgments}
This work was supported in part by U.S. Department of Energy grant No. DE-FG02-04ER41286.

\bibliography{small_instantons}
\bibliographystyle{JHEP}

\end{document}